\documentclass[reprint,pra,longbibliography,superscriptaddress,amsmath,amssymb,aps]{revtex4-2}

\usepackage{amssymb}

\usepackage{graphicx}
\usepackage{amsmath,amssymb,amsfonts}
\usepackage{amsthm}
\usepackage{mathrsfs}
\usepackage{siunitx}
\usepackage{chemformula}
\usepackage[T1]{fontenc}

\let\vaccent=\v 
\renewcommand{\v}[1]{\ensuremath{\mathbf{#1}}} 
\newcommand{\gv}[1]{\ensuremath{\mbox{\boldmath$ #1 $}}} 
\newcommand{\ket}[1]{\left| #1 \right>} 
\newcommand{\abs}[1]{\left| #1 \right|} 

\begin{document}

\title[Article Title]{Nuclear Spin Engineering for Quantum Information Science}

\author{Jonathan C. Marcks}
\altaffiliation{Corresponding authors: Jonathan C. Marcks, jmarcks@anl.gov; David D. Awschalom, awsch@uchicago.edu}
\affiliation{Q-NEXT, Argonne National Laboratory, Lemont, Illinois, 60439, United States}
\affiliation{Materials Science Division, Argonne National Laboratory, Lemont, Illinois, 60439, United States}
\affiliation{Pritzker School of Molecular Engineering, University of Chicago, Chicago, Illinois, 60637, United States}

\author{Benjamin Pingault}
\affiliation{Materials Science Division, Argonne National Laboratory, Lemont, Illinois, 60439, United States}
\affiliation{Q-NEXT, Argonne National Laboratory, Lemont, Illinois, 60439, United States}
\affiliation{Pritzker School of Molecular Engineering, University of Chicago, Chicago, Illinois, 60637, United States}

\author{Jiefei Zhang}
\affiliation{Materials Science Division, Argonne National Laboratory, Lemont, Illinois, 60439, United States}
\affiliation{Q-NEXT, Argonne National Laboratory, Lemont, Illinois, 60439, United States}

\author{Cyrus Zeledon}
\affiliation{Pritzker School of Molecular Engineering, University of Chicago, Chicago, Illinois, 60637, United States}

\author{F. Joseph Heremans}
\affiliation{Q-NEXT, Argonne National Laboratory, Lemont, Illinois, 60439, United States}
\affiliation{Pritzker School of Molecular Engineering, University of Chicago, Chicago, Illinois, 60637, United States}
\affiliation{Materials Science Division, Argonne National Laboratory, Lemont, Illinois, 60439, United States}

\author{David D. Awschalom}
\altaffiliation{Corresponding authors: Jonathan C. Marcks, jmarcks@anl.gov; David D. Awschalom, awsch@uchicago.edu}
\affiliation{Pritzker School of Molecular Engineering, University of Chicago, Chicago, Illinois, 60637, United States}
\affiliation{Q-NEXT, Argonne National Laboratory, Lemont, Illinois, 60439, United States}
\affiliation{Materials Science Division, Argonne National Laboratory, Lemont, Illinois, 60439, United States}
\affiliation{Department of Physics, University of Chicago, Chicago, Illinois, 60637, United States}

\begin{abstract}
Semiconductors are the backbone of modern technology, garnering decades of investment in high quality materials and devices. Electron spin systems in semiconductors, including atomic defects and quantum dots, have been demonstrated in the last two decades to host quantum coherent spin qubits, often with coherent spin-photon interfaces and proximal nuclear spins. These systems are at the center of developing quantum technology. However, new material challenges arise when considering the isotopic composition of host and qubit systems. The isotopic composition governs the nature and concentration of nuclear spins, which naturally occur in leading host materials. These spins generate magnetic noise---detrimental to qubit coherence---but also show promise as local quantum memories and processors, necessitating careful engineering dependent on the targeted application. Reviewing recent experimental and theoretical progress towards understanding local nuclear spin environments in semiconductors, we show this aspect of material engineering as critical to quantum information technology.

\end{abstract}

\maketitle

\section{Introduction}
\label{sec:introduction}
Information technology is a defining feature of the modern world, enabling global communication networks and advanced computing for commercial, scientific, defense, and every-day applications. These classical technologies, where complex device physics is abstracted into the information bit, require an understanding of how underlying material properties, such as a semiconductor bandgap, affect device operation~\cite{kittel_introduction_2005}. Quantum mechanics, which explains condensed matter phenomena relevant for classical technologies~\cite{wilson_theory_1931}, also predicts coherent effects---coherent superpositions, entanglement, quantum interference~\cite{sakurai_modern_1994}---that are observable and controllable in systems sufficiently isolated from their environment~\cite{bell_problem_1966,nielsen_quantum_2010}. Initially studied in atomic systems~\cite{cirac_quantum_1995,monroe_demonstration_1995}, these effects are now readily studied in solid state systems, and are core to the emerging field of quantum information science and technology (QIST)~\cite{awschalom_quantum_2018,kjaergaard_superconducting_2020,burkard_semiconductor_2023}. As with classical technology, the relevant properties of a solid state quantum bit, or qubit---coherence, control fidelity, coupling to an external system---are dictated not only by device design but also by the underlying material~\cite{de_leon_materials_2021,wolfowicz_quantum_2021}.

Notably, the isotope of the constituent atoms of a material changes not just their atomic mass but their nuclear spin. Generally irrelevant for classical technologies, nuclear spins in solid state quantum systems constitute both a dominant source of potentially detrimental magnetic noise~\cite{witzel_quantum_2005,childress_coherent_2006,chirolli_decoherence_2008,mizuochi_coherence_2009,abe_electron_2010,zhao_decoherence_2012} as well as a platform for highly coherent qubits~\cite{gurudev_dutt_quantum_2007,pla_high-fidelity_2013,muhonen_storing_2014,wolfowicz_quantum_2021}. While early demonstrations of nuclear magnetic resonance (NMR) quantum computing were performed on ensembles of nuclear spins in molecules~\cite{jones_implementation_1998,chuang_bulk_1998,chuang_experimental_1998}, presently the engineering of nuclear spins for quantum applications mainly arises in the context of electron spins in semiconductors~\cite{kanai_generalized_2022,wolfowicz_quantum_2021,burkard_semiconductor_2023}.

\begin{figure}
    \centering
    \includegraphics{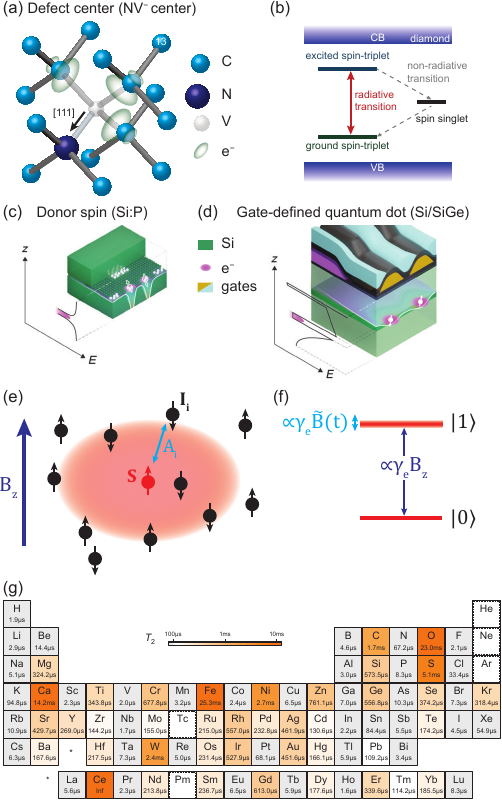}
    \caption{\bf Electron-nuclear spin systems. \rm (a) Nitrogen-vacancy (NV) center in diamond, a prototypical spin defect color center. (b) A simplified level structure of the NV center. (c) \ch{^{31}P} donor spin qubit in Si. (d) Gate-defined quantum dot spin qubits in a Si/SiGe heterostructure. The graphs in (c) and (d) plot the confining potential. Reprinted with permission from Ref.~\cite{burkard_semiconductor_2023}. Copyright 2023 by the American Physical Society. (e) A central spin $\v{S}$ in a spin bath consisting of nuclear spins $\v{I}_i$ in an external magnetic field $B_z$. Shaded area shows wavefunction extent. (f) Qubit level diagram with coupling to fluctuating magnetic field. (g) Periodic table for quantum coherence showing CCE-calculated coherence times given natural isotope abundance. Adapted from Ref.~\cite{kanai_generalized_2022}.}
    \label{fig:systems}
\end{figure}

Electron spins may be hosted in color centers with electron spin-photon interfaces~\cite{doherty_nitrogen-vacancy_2013,atature_material_2018,awschalom_quantum_2018}, donor atoms~\cite{kane_silicon-based_1998,pla_high-fidelity_2013,burkard_semiconductor_2023}, or gate-defined quantum dots (QDs)~\cite{petta_coherent_2005,burkard_semiconductor_2023}. Color centers consist of atomic vacancies, substitutions, interstitials, or some combination thereof in the crystal lattice, such as the nitrogen vacancy (NV) center in diamond~\cite{doherty_nitrogen-vacancy_2013} shown in Fig.~\ref{fig:systems}(a). They possess an atom-like level structure that allows optical control and measurement of the color center's electron spin~\cite{jelezko_read-out_2004,robledo_high-fidelity_2011,nagy_high-fidelity_2019} [Fig.~\ref{fig:systems}(b)]. Color centers can also rely on the electronic structure of impurity atoms, as with rare-earth ions~\cite{zhong_nanophotonic_2015,dibos_atomic_2018,kindem_control_2020,raha_optical_2020}. Experiments over the past two decades have established single color centers and ensembles of color centers as a new class of sensor~\cite{maze_nanoscale_2008,taylor_high-sensitivity_2008,barry_sensitivity_2020} and networking~\cite{gao_coherent_2015,reiserer_robust_2016,ruf_quantum_2021,bradley_robust_2022,bushmakin_two-photon_2024} building blocks. Electrons confined by electronic potentials, arising from dopants~\cite{fuechsle_single-atom_2012,pla_high-fidelity_2013} [Fig.~\ref{fig:systems}(c)] or the gating of an electronic gas in a semiconductor heterostructure~\cite{kouwenhoven_excitation_1997,eriksson_spin-based_2004,petta_coherent_2005,burkard_semiconductor_2023}, forming QDs [Fig.~\ref{fig:systems}(d)], are promising for quantum computation~\cite{kane_silicon-based_1998,hill_surface_2015,vandersypen_interfacing_2017}. These systems are measured electrically, with spin-to-charge conversion schemes~\cite{elzerman_single-shot_2004,lai_pauli_2011}.

In all cases, electron spins may interact with nuclear spins both inherent to the qubit system~\cite{felton_hyperfine_2009,sangtawesin_hyperfine-enhanced_2016} and dispersed in the host material~\cite{maletinsky_dynamics_2007,yang_quantum_2008,chekhovich_nuclear_2013,abobeih_fault-tolerant_2022,goldblatt_sensing_2024}. Further development of quantum technologies will benefit from engineering the isotopic composition and density of nuclear spins~\cite{bourassa_entanglement_2020,neyens_probing_2024}, as well as, in tandem, coherent coupling between electron spins and nuclear spins~\cite{abobeih_fault-tolerant_2022}, and between nuclear spins~\cite{thorvaldson_grovers_2024}. In this perspective, we review the critical role nuclear spins play in emerging quantum technologies and demonstrate the advantages of isotopically engineering crystal environments.

\section{Electron-nuclear spin interactions}
Many quantum applications already use nearby nuclear spins as quantum memories for sensing~\cite{zaiser_enhancing_2016,rosskopf_quantum_2017,pfender_nonvolatile_2017,arunkumar_quantum_2023} and networking~\cite{pompili_realization_2021,knaut_entanglement_2024}, and as qubits~\cite{reiner_high-fidelity_2024,thorvaldson_grovers_2024}. Recent work has also shown that nuclear spins can be harnessed for quantum simulation~\cite{randall_many-bodylocalized_2021} and the realization of logical qubits~\cite{abobeih_fault-tolerant_2022}. In general, nuclear spins may cause decoherence of central electron spins~\cite{yao_theory_2006,childress_coherent_2006,mikkelsen_optically_2007,yang_quantum_2008,maze_electron_2008,ma_uncovering_2014,neyens_probing_2024}. A typical system of one electronic spin in an environment of many nuclear spins [Fig.~\ref{fig:systems}(e)] is described by the Hamiltonian of the electron system, $\hat H_S$, the nuclear spins, $\hat H_I$, and the interactions, $\hat H_{SI}$~\cite{onizhuk_decoherence_2024}

\begin{eqnarray}
    \label{eq:H}
    \hat H = \hat H_S + \hat H_I + \hat H_{SI}\nonumber\\
    \label{eq:HS}
    \hat H_S = \v S\cdot \v D \cdot \v S + \v B\cdot \gv\gamma_e\cdot\v S\nonumber\\
    \label{eq:HB}
    \hat H_I = \sum_i \left(\v I_i\cdot \v P_i\cdot \v I_i + \v B\cdot\gv\gamma_{n,i}\cdot\v I_i\right) \nonumber\\
    + \sum_{i<j}\v I_i\cdot \v J_{ij}\cdot \v I_j\nonumber\\
    \label{eq:HSB}
    \hat H_{SI} = \sum_i \v S\cdot \v A_{i} \cdot \v I_i
\end{eqnarray}
where $\v S=(\hat S_x,\hat S_y,\hat S_z)$ is the electron spin operator, $\v D$ describes the electron self-interaction (crystal field splitting, non-zero for $S>1/2$), $\v B$ is the external magnetic field vector, $\gv\gamma$ is the electron gyromagnetic ratio tensor, which in general may be anisotropic, $\v I_i=(\hat I_{x,i}, \hat I_{y,i}, \hat I_{z,i})$ is the nuclear spin operator, $\v P_i$ is the nuclear self-interaction (non-zero for $I>1/2$), $\gv\gamma_{n,i}$ is the nuclear gyromagnetic ratio, $\v J_{ij}$ is the nuclear-nuclear interaction, and $\v A_{i}$ is the electron-nuclear interaction tensor. $i$ refers to the $i^{\text{th}}$ nuclear spin. $\v A$ contains Fermi contact, dipolar (hyperfine), and higher-order spin-orbit contributions.

During some quantum operation, the electron spin will typically start in a coherent superposition of logical $\ket 0$ and $\ket 1$ states, $\ket{\psi_e}=\alpha\ket 0+\beta\ket 1$, which can equivalently be written with the density matrix $\rho_e=\left(\begin{array}{cc}\abs{\alpha}^2&\alpha^*\beta\\\alpha\beta^*&\abs{\beta}^2\end{array}\right)$, where we can define $W=\alpha\beta^*$ as the coherence. As a result of the electron-nuclear spin-spin interactions, nuclear spin flips and nuclear-nuclear flip-flops in general detune the energy of the electron spin [Fig.~\ref{fig:systems}(f)], leading to coherence decay~\cite{degen_quantum_2017} of $W(t)=\exp[-\chi(t)]$, assuming nuclear spins dominate decoherence, where the functional form of $\chi(t)$ depends on the nuclear spin behavior. If the coupling between the electron spin and a nuclear spin $A_i$ is large enough relative to the decoherence, the electron spin may be used to coherently control the nuclear spin~\cite{bourassa_entanglement_2020}, enabling, e.g., nuclear spin quantum memories. The nuclear spin density will impact both decoherence and the coherent coupling~\cite{yang_quantum_2008,bourassa_entanglement_2020}.

\section{Enhancing quantum coherence}

While crystal defects and bound states exist in all solid state systems~\cite{pantelides_electronic_1978}, suitable qubit host materials must have (i) a wide bandgap ($\geq\SI{1}{\eV}$) to host a qubit's atom-like energy level structure, (ii) elements with a stable nuclear spin-free isotope to enhance the qubit coherence time $T_2$ and enable isotopic engineering of the environment, and ideally (iii) low spin-orbit coupling to extend the qubit lifetime $T_1$~\cite{weber_quantum_2010}. Silicon and carbon-based semiconductors thus emerge as prime candidates for spin qubit hosts. Recent theoretical work has predicted that many oxides may also host long-lived coherent spin qubits~\cite{kanai_generalized_2022}. The ``periodic table of coherence'' from Ref.~\cite{kanai_generalized_2022} is reproduced in Fig.~\ref{fig:systems}(g), demonstrating a range of coherence time predictions dependent on the naturally abundant nuclear species. Table~\ref{tbl:SiCisotopes} lists C, Si, N, and O isotope abundance, nuclear spin, and achievable purification level. The monoatomic diamond and silicon, as well as the diatomic silicon carbide (SiC), host electron spin systems that are under widespread study. These materials are studied for electronics applications, leading to a large body of synthesis knowledge. Other wide-bandgap semiconductors, such as GaN and BN, are less ideal qubit hosts due to the lack of nuclear spin-free isotopes.

Cluster correlation expansion (CCE) calculations of various semiconductor host materials reveal that beyond the abundance of nuclear spins, the crystal structure and density, as well as the relative concentration of different nuclear spin species, will also impact the electron spin coherence~\cite{kanai_generalized_2022} [Fig.~\ref{fig:coherence}(a-c)]. This reveals, for example, that while the relatively low nuclear spin concentration in diamond extends coherence, the dense packing of carbon atoms decreases it~\cite{seo_quantum_2016}.

\begin{table}
\caption{\label{tbl:SiCisotopes}Stable isotopes of carbon, silicon, and nitrogen with their nuclear spins, natural purity, and typical achieved purities. Purity levels are only quoted for the nuclear spin-containing isotopes of C and Si.}
    \begin{tabular}{|c|c|c|c|}
    \hline
        isotope & nuclear spin & nat. abundance & typ. purity\\
        \hline
        \ch{^{12}C} & 0 & 98.9\% &\\
        \ch{^{13}C} & 1/2 & 1.1\% & \SI{10}{ppm}\\ 
        \ch{^{28}Si} & 0 & 92.2\% &\\
        \ch{^{29}Si} & 1/2 & 4.7\% & \SIrange{50}{800}{ppm}\\
        \ch{^{30}Si} & 0 & 3.1\% &\\
        \ch{^{14}N} & 1 & 99.6\% &\\
        \ch{^{15}N} & 1/2 & 0.4\% &\\
        \ch{^{16}O} & 0 & 99.8\% &\\
        \ch{^{17}O} & 5/2 & 0.04\% &\\
        \ch{^{18}O} & 0 & .2\% &\\
        \hline
        
    \end{tabular}
\end{table}

\subsection{Group-IV materials}
\begin{figure}
    \centering
    \includegraphics{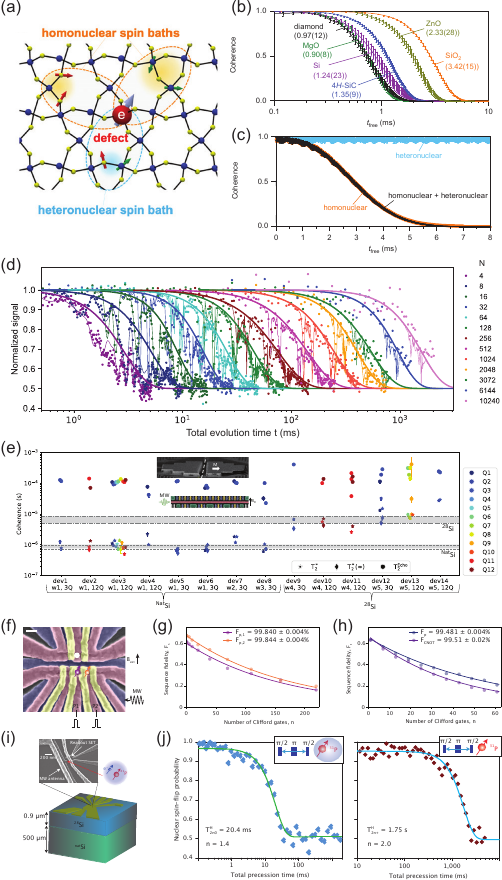}
    \caption{\bf Coherence engineering in semiconductors. \rm (a) Schematic of CCE calculations for an electron spin in both homonuclear and heteronuclear environments. (b) CCE-calculated Hahn echo coherence for various defect host materials. (c) CCE-calculated Hahn echo coherence in \ch{SiO2} for differing nuclear spin baths. Adapted from Ref.~\cite{kanai_generalized_2022}. (d) NV center electron spin coherence under dynamical decoupling with $N$ pulses in a natural-abundance diamond. Adapted from Ref.~\cite{abobeih_one-second_2018}. (e) Coherence times of 39 gate-defined quantum dot spin qubits formed across 14 Si/SiGe devices of varying geometry and isotopic purity (SEM inset). Adapted from Ref.~\cite{neyens_probing_2024}. (f) SEM micrograph of a two-qubit gate-defined quantum dot system. Scale bar is \SI{100}{\nano\meter}. (g,h) One- and two-qubit gate fidelities measured with Clifford-based randomized benchmarking. Adapted from Ref.~\cite{noiri_fast_2022}, reprinted with permission from Springer Nature. (i) SEM micrograph and schematic of \ch{^{31}P} donor system in isotopically purified Si. (j) \ch{^{31}P} nuclear spin Hahn echo coherence with (left) and without (right) the electron present. Adapted from Ref.~\cite{muhonen_storing_2014}, reprinted with permission from Springer Nature.}
    \label{fig:coherence}
\end{figure}

Among diamond's large number of atomic impurities, the nitrogen-vacancy (NV) center and the silicon-vacancy (SiV) center have so far been the most commonly used for quantum applications. The coherence time of their electronic spin can reach \SI{1}{\second}~\cite{abobeih_one-second_2018} [Fig.~\ref{fig:coherence}(d)]. The dephasing due to fluctuations in the spin bath of the remaining 1.1\% of \ch{^{13}C} spin-1/2 can be mitigated by isotopic purification of the diamond matrix~\cite{mizuochi_coherence_2009,balasubramanian_ultralong_2009,maurer_room-temperature_2012,ishikawa_optical_2012,yamamoto_extending_2013}. However, nuclear spins themselves can display coherence times from seconds to minutes~\cite{bartling_entanglement_2022}, such that \ch{^{13}C} nuclear spins are used as long coherence quantum registers~\cite{bradley_ten-qubit_2019}, including for entanglement~\cite{pompili_realization_2021,stas_robust_2022,knaut_entanglement_2024} and teleportation protocols~\cite{hermans_qubit_2022}. Furthermore, dynamical decoupling pulse sequences performed on the color center spin can extend the coherence time beyond what is obtained in isotopically purified diamonds~\cite{abobeih_one-second_2018}.

Silicon carbide is also host to many optically active spins~\cite{atature_material_2018,awschalom_quantum_2018,castelletto_silicon_2020}, the most widely studied of which are the neutral divacancy (VV) and the negatively charged silicon vacancy (V$_{Si}$) in the 4H polytype~\cite{mizuochi_continuous-wave_2002,koehl_room_2011,baranov_silicon_2011,soltamov_room_2012,falk_electrically_2014,kraus_room-temperature_2014,widmann_coherent_2015,christle_isolated_2015}. With natural nuclear spin abundance in SiC (Tbl.~\ref{tbl:SiCisotopes}), the electron spin dephasing times reach $\sim\SI{1}{\mu\second}$~\cite{christle_isolated_2015,nagy_quantum_2018} ($\sim\SI{100}{\mu\second}$ for basal VV operating at a zero first-order Zeeman (ZEFOZ) transition~\cite{miao_universal_2020}), and dynamical decoupling can further extend their coherence time to tens of milliseconds for both VV~\cite{bourassa_entanglement_2020} and V$_{Si}$~\cite{simin_locking_2017}.

These coherence times are on par with those observed in diamond, even though SiC displays a higher natural nuclear spin content from the \ch{^{29}Si} and \ch{^{13}C} isotopes (with natural abundance of $4.7\%$ and $1.1\%$ respectively). However, the gyromagnetic ratio mismatch between the nuclear spins of these two isotopes effectively results in two independent spin baths with longer atomic bonds than in diamond \cite{yang_electron_2014,christle_isolated_2015,widmann_coherent_2015,seo_quantum_2016}. As in diamond, isotopic engineering can lead to increased coherence times for both VV \cite{bourassa_entanglement_2020} and V$_{Si}$ \cite{nagy_quantum_2018}, with a value of 5 seconds achieved in combination with dynamical decoupling \cite{anderson_five-second_2022}.

In silicon, isotopic purification has been pushed to the greatest extent through the Avogadro project~\cite{becker_enrichment_2010} with an isotopic abundance of $^{28}$Si reaching 99.995$\%$ ($[\ch{^{29}Si}]=\SI{50}{ppm}$). This has been leveraged to improve the coherence properties of the T center electron spin in silicon to \SI{2.1}{\milli\second} for the electron and \SI{1.1}{\second} for the associated hydrogen nuclear spin using a Hahn echo sequence~\cite{bergeron_silicon-integrated_2020}.

\ch{^{29}Si} nuclear spins may also limit qubit coherence and gate fidelity in donor spin and gate-defined quantum dots, where the electronic wavefunction extends over a nm-scale~\cite{burkard_semiconductor_2023}. The wavefunction may overlap with many nuclear spins, leading to a dominant Fermi contact interaction, depending on the nuclear spin density~\cite{witzel_electron_2010}. As the electron confinement is electrically controlled, charge noise, ubiquitous in semiconductors~\cite{van_der_ziel_unified_1988}, can perturb the location and size of the electron wavefunction, and thus the interactions with the nuclear spin bath. This effectively transduces electrical noise into magnetic noise, in addition to the intrinsic magnetic noise arising from nuclear spin fluctuations.

In gate-defined QDs, isotopic purification has been demonstrated to improve coherence times by an order of magnitude in industrially fabricated devices, where a clear improvement across many devices is observed~\cite{neyens_probing_2024} [Fig.~\ref{fig:coherence}(e)]. Removing the background nuclear spin noise enables one- and two-qubit gates that are compatible with fault-tolerant quantum computing~\cite{veldhorst_addressable_2014,noiri_fast_2022} [Fig.~\ref{fig:coherence}(f-h)]. In naturally abundant material, dynamic nuclear polarization may enable nuclear memories and improved spin coherence~\cite{appel_many-body_2025,cai_formation_2025}, and real-time feedback may mitigate the effects of magnetic noise~\cite{berritta_real-time_2024,berritta_physics-informed_2024}.

The Kane proposal for a quantum computer~\cite{kane_silicon-based_1998} utilizes the \ch{^{31}P} nuclear spin as the qubit, such that beyond the effect of background nuclear spin noise on the electron spin state, which may affect control of the nuclear spin, the relevant property is the nuclear spin coherence time. By ionizing the donor---removing the electron but leaving the coherent nuclear spin state intact---the coherence time has been shown to improve by two to three orders of magnitude~\cite{muhonen_storing_2014} [Fig.~\ref{fig:coherence}(i,j)].

A complete and on-going overview of advances in QD coherence and gate fidelity may be found in Ref.~\cite{stano_review_2022}. Presently, the longest reported coherence times (with no advanced driving protocols) and highest reported single qubit gate fidelities are $T_2^*=\SI{235}{\mu\second}$~\cite{hansen_implementation_2022}, $T_{2,hahn}=\SI{1.2}{\milli\second}$~\cite{veldhorst_addressable_2014}, $\mathcal F=99.957\%$~\cite{yang_silicon_2019} for \ch{^{28}Si}/\ch{SiO2} quantum dots; $T_2^*=\SI{20}{\mu\second}$~\cite{struck_low-frequency_2020}, $T_{2,hahn}=\SI{1}{\milli\second}$~\cite{kerckhoff_magnetic_2021}, $\mathcal F=99.975\%$~\cite{blumoff_fast_2022} for \ch{^{28}Si}/\ch{SiGe} quantum dots; and $T_2^*=\SI{270}{\mu\second}$~\cite{muhonen_storing_2014}, $T_{2,hahn}=\SI{1.73}{\milli\second}$~\cite{madzik_controllable_2020}, $\mathcal F=99.942\%$~\cite{dehollain_optimization_2016} for \ch{^{28}Si}:\ch{^{31}P} donor electron spins. We note that the highest single qubit gate fidelity, $\mathcal F=99.992\%$~\cite{lawrie_simultaneous_2023} in quantum dot/donor systems has been observed in Ge/SiGe (Ge quantum well), a newer material platform.

\subsection{Rare-earth oxides}
\begin{figure}
    \centering
    \includegraphics{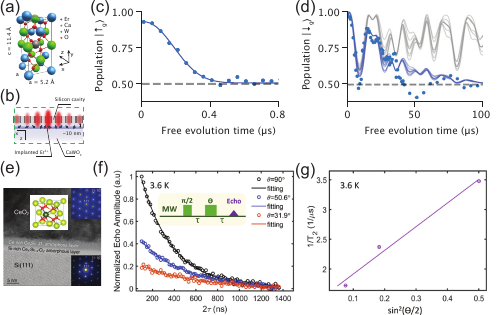}
    \caption{\bf Rare earth ions in oxide hosts. \rm (a) \ch{Er^{3+}} ion in \ch{CaWO4} host. (b) A Si nanophotonic cavity enhances photonic coupling to dim \ch{Er^{3+}} emitters. (c) \ch{Er^{3+}} electron spin Ramsey coherence. (d) \ch{Er^{3+}} electron spin Hahn echo coherence, overlaid with CCE-calculated coherence from nuclear spins without (gray) and with (blue) extra overall decay. Adapted from Ref.~\cite{ourari_indistinguishable_2023}, reprinted with permission from Springer Nature. (f) Epitaxial growth of \ch{CeO2} on Si(111), with \ch{Er^{3+}} ion in \ch{CeO2} host inset. Adapted from Ref.~\cite{grant_optical_2024}. (f,g) Generalized Hahn echo coherence of \ch{Er^{3+}} electron spin, revealing instantaneous diffusion effects. Adapted from Ref.~\cite{zhang_optical_2024}.}
    \label{fig:REI}
\end{figure}

Quantum networks will require coherent interfaces between optical photons and other long-lived degrees of freedom or processing units. Individually addressable solid-state defects integrated with nanophotonic devices are of particular interest. Typical color centers in wide-bandgap semiconductors (e.g., diamond and SiC discussed above) have atomic transitions covering \SIrange{400}{1100}{\nano\meter}, outside the low-loss telecom band in standard optical fibers ($\SI{1.5}{\mu\meter}$) for long-distance transmission. The rare earth ion (REI) Er$^{3+}$ provides a direct spin-photon interface in the telecom band and has been a recent research focus. There are challenges with exploiting individual REIs as single photon sources and quantum memories coming from their low photon emission rates and correspondingly increased sensitivity to decoherence. Recent work has demonstrated incorporating REIs into nanophotonic cavities at the targeted ion transition frequency to enhance the emission of choice with a factor of $\approx 1000$ to overcome the slow photon emission bottleneck ~\cite{dibos_atomic_2018,uysal_spin-photon_2024}. 

The path towards mediation of decoherence lies in the careful selection and engineering of host materials. As a Kramer's ion, the intrinsic non-zero electronic magnetic moment of Er$^{3+}$ imposes an intrinsic limit on the electronic spin relaxation and, therefore, the spin coherence. In addition, the presence of fluctuating magnetic field noise induced by the intrinsic electronic and nuclear spins in host materials further reduces the spin coherence. Therefore, finding suitable host materials is key to improving spin properties for network applications. To this end, a high-throughput simulation using cluster correlation expansion (CCE) has been implemented to predict spin coherence from the perspective of nuclear spins and screen over 12,000 hosts to address the question of suitable hosts for targeted applications~\cite{kanai_generalized_2022}.

Er$^{3+}$ has been largely explored in bulk crystals such as \ch{Y_2SO_5} (YSO), \ch{Y2O3}, or \ch{YVO4} (YVO). Electron spin coherence times of $\SI{5.6}{\mu\second}$ have been reported for Er$^{3+}$ ions in YSO~\cite{probst_microwave_2015}, at \SI{30}{\milli\kelvin}, limited by yttrium’s high first-order magnetic sensitivity reaching up to \SI{200}{\giga\hertz\per\tesla}. For this reason, most demonstrations of long spin coherence in yttrium-based matrices have relied on magnetically insensitive electron-spin transitions such as ZEFOZ, or clock, transitions. A coherence time of up to \SI{1}{\milli\second} has been reached in \ch{Y2O3}~\cite{gupta_robust_2023}. To reduce nuclear spin content, \ch{CaWO4} has been examined [Fig.~\ref{fig:REI}(a)]. Promisingly, an electron spin coherence time of \SI{23}{\milli\second} has been reported for sub-ppb (parts-per-billion) Er$^{3+}$ at \SI{10}{\milli\kelvin} in \ch{CaWO4}, which has a sparse nuclear spin environment (only 14\% natural abundance of the spin-$\frac{1}{2}$ \ch{^{183}W} isotope) contributing to the spin noise~\cite{le_dantec_twenty-threemillisecond_2021}, consistent with the CCE prediction. However, in systems seeking to leverage the optical spin-photon interface in fabricated structures [Fig.~\ref{fig:REI}(b)], comparison to CCE calculations suggests there are non-nuclear limits to coherence~\cite{ourari_indistinguishable_2023} [Fig.~\ref{fig:REI}(c-d)].

Regardless, finding host materials with a low natural abundance of isotopes with non-zero nuclear spins, circumventing the need for isotopic purification, is a viable pathway to long-coherence REI systems. Cerium dioxide (\ch{CeO2}) [Fig.~\ref{fig:REI}(e)], with cerium contributing zero nuclear spin and oxygen carrying only 0.04\%\ (\ch{^{17}O}), is a promising potential host for spin qubits, with a theoretically predicted coherence time up to \SI{47}{\milli\second}~\cite{kanai_generalized_2022}. The compatibility of \ch{CeO2} with silicon and its engineerable piezoelectric property also make the material a good candidate for quantum opto-electronic devices. Recently, the molecular beam epitaxy (MBE) of single-crystal Er-doped \ch{CeO2} films on Si(111) substrates was demonstrated~\cite{grant_optical_2024} [Fig.~\ref{fig:REI}(e)], alongside an exploration of the Er$^{3+}$ spin coherence in the system~\cite{zhang_optical_2024}. A Hahn echo spin coherence of \SIrange{250}{500}{\nano\second} is observed at \SI{3.6}{\kelvin} in samples with high Er concentration (\SIrange{2}{5}{ppm}) [Fig.~\ref{fig:REI}(f,g)]. The low nuclear spin bath concentration in \ch{CeO2} enables the direct observation and measurement of Hahn echo at \SI{4}{\kelvin} which is not observable in YSO, \ch{YVO4}, and \ch{CaWO4}~\cite{raha_optical_2020}, pointing to the potential ability to host long-lived spins.   

Separately, there is also the path of isotopic purification to reduce nuclear spins. Er$^{3+}$ in silicon is a well explored platform showing narrow optical transition~\cite{gritsch_narrow_2022} and spin coherence up to $\SI{48}{\mu\second}$~\cite{gritsch_optical_2025}, limited by interaction with \ch{^{29}Si}. With isotopic purification ([\ch{^{29}Si}]$<$0.01\%) Hahn echo electron spin coherence times of \SI{0.8}{\milli\second} and \SI{1.2}{\milli\second} have been demonstrated for two different Er sites at sub-kelvin temperature~\cite{berkman_millisecond_2023}. 

These works point to the need and promise of host engineering to reduce nuclear spin concentration and improve spin coherence. With the nuclear spin effect nominally removed, one would also need to consider other limiting factors within specific material platforms, such as dopant impurities in the host, crystallographic distortions, charging, and strain. Of particular note is the difficulty in separating rare-earth ions from each other~\cite{opare_comparative_2021,chen_recent_2022}, often resulting in undesired rare-earth impurities within rare-earth oxides.  This in turn motivates the need for higher purity precursor sources to reduce the background dopant contaminants in the host crystal. Systematic investigation of these critical factors beyond nuclear spins are needed for the continued development of quantum host materials and devices for various quantum applications. 

\section{Nuclear spin memories and processors}
\begin{figure*}
    \centering
    \includegraphics[width=\textwidth]{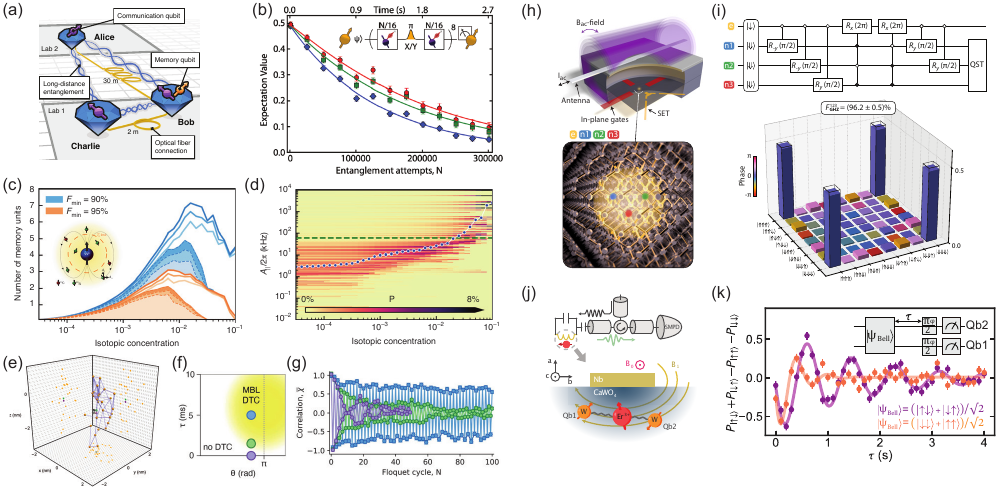}
    \caption{\bf Nuclear memories and processors. \rm (a) Three-node quantum network of NV centers in diamond with a nuclear spin memory. Adapted from Ref.~\cite{pompili_realization_2021}, reprinted with permission from AAAS. (b) Nuclear spin expectation value averaged over six cardinal states for $N$ electron spin remote entanglement attempts, shown for three post-selection criteria. Adapted from Ref.~\cite{bradley_robust_2022}. (c) Average number of nuclear memories for a VV defect in SiC (inset) versus nuclear spin-containing isotope concentration ([\ch{^{13}C}]=[\ch{^{29}Si}]). Dashed line areas refer only to distributions where all hyperfine couplings are $<2\pi\times\SI{60}{kHz}$. (d) Hyperfine value distributions for isotope concentrations. Green line is at $2\pi\times\SI{60}{kHz}$. Adapted from Ref.~\cite{bourassa_entanglement_2020}, reprinted with permission from Springer Nature. (e) 3D structure of \ch{^{13}C} nuclear spins surrounding a NV center in diamond, measured with a double-resonance 3D spectroscopy technique. A spin chain is identified from this structure for quantum simulations. Adapted from Ref.~\cite{abobeih_atomic-scale_2019}, reprinted with permission from Springer Nature. (f-g) Many-body localized (MBL) discrete time-crystal (DTC) phase transition measured on a nuclear spin quantum simulator. Adapted from Ref.~\cite{randall_many-bodylocalized_2021}, reprinted with permission from AAAS. (h) Schematic of electrically detected \ch{^{31}P} nuclear spin processor in Si. (i) Generation of a three-nuclear spin GHZ state in Si. Adapted from Ref.~\cite{thorvaldson_grovers_2024}. (j) Schematic of a microwave photon-detected \ch{Er^{3+}} electron spin coupled to naturally occurring \ch{^{183}W} nuclear spins in \ch{CaWO4}. (k) Ramsey coherence of \ch{^{183}W} nuclear spin Bell states. Adapted from Ref.~\cite{osullivan_individual_2024}.}
    \label{fig:processors}
\end{figure*}

When electron spins are interfaced with individual nuclear spins, applications of hybrid electron-nuclear systems may be realized. Nuclear gyromagnetic ratios are around three orders of magnitude smaller than the electronic gyromagnetic ratio, leading to significantly longer nuclear coherence times, while at the same time leading to a smaller coupling to control signals. Nuclear spins also generally lack the optical or electronic interfaces that enable control and measurement of the defect, donor, and QD systems discussed in this review. However, electron spins interfaced with proximal nuclear spins can control and measure nuclear spin states, enabling nuclear quantum memory, simulation, and computing platforms.

It is generally agreed that so-called quantum repeaters are necessary to distribute quantum information over many-km distances, a requirement for quantum networks~\cite{wehner_quantum_2018,pettit_perspective_2023,azuma_quantum_2023}. Quantum repeater nodes require a coherent interface between a flying qubit (photon) and stationary qubit (e.g., electron spin), afforded by a color center, as in Fig.~\ref{fig:processors}(a), or neutral atom qubit~\cite{covey_quantum_2023} to generate spin-photon entanglement~\cite{uysal_spin-photon_2024,ruskuc_scalable_2024}. Not every attempt to generate entanglement is successful, necessitating local quantum memories to store quantum information between entanglement attempts. Nuclear spins in the host matrix may be controlled by a color center and act as a quantum memory~\cite{bourassa_entanglement_2020}. These quantum memories must be long-lived not only relative to operating timescales of the network, but to the operations performed on the local electron spin qubit, which may be improved through dynamical control and isotopic purification~\cite{bradley_robust_2022}. \ch{^{13}C} nuclear spins have been shown to be robust to over $10^5$ NV center entangling operations in a quantum network~\cite{bradley_robust_2022} [Fig.~\ref{fig:processors}(b)]. However, rather than maximally purifying a host material to improve central spin coherence~\cite{bradley_robust_2022}, nuclear spin quantum memories may require an intermediate nuclear spin concentration that maximizes the number of accessible nuclear spins. Ref.~\cite{bourassa_entanglement_2020} demonstrated this for the case of SiC [Fig.~\ref{fig:processors}(c-d)], showing that lower nuclear spin concentrations can lead to weaker-coupled but better-resolved memories. This may ultimately lead to a more powerful quantum repeater node, demonstrating the importance of engineering the isotopic environment for specific applications rather than focusing on bare coherence.

The control and measurement of individual nuclear spins shift the focus from the electron spin, in a nuclear spin bath, to a system of coherent nuclear spins, with an electron spin to measure or to mediate interactions. Recent work has demonstrated the spatial mapping of up to 50 nuclear spins in the environment of an NV center~\cite{zopes_three-dimensional_2018,abobeih_atomic-scale_2019,cujia_parallel_2022,van_de_stolpe_mapping_2024} [Fig.~\ref{fig:processors}(e)]. Despite the disorder in nuclear-nuclear coupling due to the stochastic nuclear spin locations, effective spin chains, for example, may be formed through the appropriate choice of nuclear spins in such a system. While nascent, these systems have been used as quantum simulators to study many-body-localization and time crystal physics~\cite{randall_many-bodylocalized_2021} [Fig.~\ref{fig:processors}(f-g)]. Nuclear spins have also been deployed as small-scale quantum processors with interactions mediated through a single electron spin~\cite{abobeih_fault-tolerant_2022,reiner_high-fidelity_2024,osullivan_individual_2024} and with direct electrical and magnetic driving, as in Si:Sb systems~\cite{asaad_coherent_2020,fernandez_de_fuentes_navigating_2024,yu_schrodinger_2025}. In NV centers, Si:P donors, and \ch{CaWO4}:Er systems these processors can generate many-body entangled states and implement quantum algorithms~\cite{abobeih_fault-tolerant_2022,thorvaldson_grovers_2024,osullivan_individual_2024} [Fig.~\ref{fig:processors}(h-i)]. In particular, the recent advances in microwave photon detection of Er electron spins demonstrate a promising platform for electron and nuclear spins in oxides that circumvents the low optical emission rate of Er ions~\cite{le_dantec_twenty-threemillisecond_2021,osullivan_individual_2024} [Fig.~\ref{fig:processors}(j-k)].

\begin{figure}
    \centering
    \includegraphics{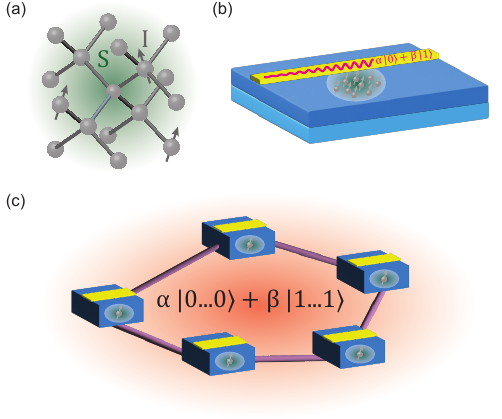}
    \caption{\bf Networked nuclear spins. \rm (a) A single electron spin interfaced with multiple nuclear spins inside a semiconductor. (b) An electron and nuclear spin-containing device with an optical, microwave, or electrical link for transmitting or measuring quantum information. (c) A network of many nuclear spin-based nodes/processors sharing a collective entangled quantum state.}
    \label{fig:network}
\end{figure}

We have reviewed how a combination of materials science and quantum engineering drives advances in coherent electron-nuclear spin systems. In Fig.~\ref{fig:network} we show the future prospect of networked nuclear spin modules, which would integrate with existing fiber optical networks and may enable quantum repeaters. Fig.~\ref{fig:network}(a) shows a group of nuclear spins in a semiconductor with an associated electron spin. The electron spin may be from a color center, a donor atom, or a quantum dot, and will likely be necessary for interfacing with the nuclear spin qubits. Fig.~\ref{fig:network}(b) has the electron-nuclear spin cluster embedded in a generic heterostructure or device with a wire for transmitting quantum states (e.g., an optical or microwave waveguide). There may be many clusters in a single chip wired together. At a larger scale, in Fig.~\ref{fig:network}(c), many devices are networked together, likely through an optical quantum network, with intermediate quantum interconnects that entangle nuclear spins and photons. These networks will host delocalized entangled states shared among nuclear spin clusters in each device, and may enable goals in the field such as distributed quantum computing and overcoming wiring input/output problems that arise from scaling individual quantum processors.

\section{Conclusion}
The study of coherent quantum systems has revealed new ways to process information at a fundamental level, leading to a growing landscape of quantum technologies. Many qubit systems, whether for computing, communication, or sensing applications, are hosted in semiconductors, where relevant properties---bandgap, spin-orbit coupling, spin coherence---depend on the atomistic details of constituent elements, such as the isotope and associated nuclear spin (or lack thereof).  In this perspective, we focused on the role that the engineering and manipulation of nuclear spins in semiconductors will play in emerging quantum technology. Group-IV semiconductors, such as diamond and Si, and various oxides are promising hosts not only for coherent electronic spin qubits, which are readily interfaced electrically and/or optically and benefit from low nuclear spin densities, but also for nuclear spin qubits, which themselves can sustain long-lived quantum states. Ultimately, the approach to engineering these nuclear spins, e.g., the degree of isotopic purification, depends on the final application, ranging from highly coherent electron spin sensors to a high number of nuclear memories. We expect nuclear spins to be an integral part of future quantum technology.

\section{Declarations}
The authors thank Nazar Delegan for useful discussions and insight into isotopic growth and Marquis McMillan for comments on the manuscript.

This work was supported by Q-NEXT, a U.S. Department of Energy Office of Science National Quantum Information Science Research Centers under Award Number DE-FOA-0002253 (J.C.M., B.P., D.D.A.), the U.S. Department of Energy, Office of Science, Basic Energy Sciences, Materials Sciences and Engineering Division through Argonne National Laboratory under Contract No. DE-AC02-06CH11357 (J.Z., F.J.H.), the Air Force Office of Scientific Research under award number FA9550-23-1-0330 (C.Z.), and Boeing through the Chicago Quantum Exchange (C.Z.).

The authors declare no conflicts of interest.

\end{document}